\documentclass{moriond}

\usepackage[utf8]{inputenc}
\usepackage[T1]{fontenc}
\usepackage[rightcaption]{sidecap}
\sidecaptionvpos{figure}{c}
\usepackage{float}
\usepackage{amsmath,amssymb}
\usepackage{dsfont}
\usepackage{hyperref}
\usepackage{graphicx}
\usepackage{enumitem}
\usepackage{mathtools}
\usepackage{slashed}
\usepackage{bbold}
\usepackage{multirow}
\usepackage{ytableau}
\usepackage{youngtab}
\usepackage{braket}
\usepackage{soul}
\usepackage{cancel}
\usepackage{xcolor}
\usepackage[normalem]{ulem}

\usepackage{lipsum}

\usepackage{colortbl}
\def\be{\begin{equation}}
\def\ee{\end{equation}}
\def\bea{\begin{eqnarray}}
\def\eea{\end{eqnarray}}

\newcommand{\Photo}{}
\begin{document}
\vspace*{4cm}

\title{Probing the neutrino mass through semileptonic decays}

\author{D. Be\v{c}irevi\'c$^a$, C. Chevallier$^a$, S. Fajfer$^{b,c}$, N. Ko\v{s}nik$^{b,c}$ and L. Pavi\v{c}ić$^c$}

\address{\small $^a$IJCLab, CNRS/IN2P3 et Universit\'e Paris-Saclay,\quad
$^b$University of Ljubljana,\quad
$^c$Jo\v{z}ef Stefan Institute}

\maketitle\abstracts{We argue that a detailed analysis of semileptonic decays can test the possibility of a massive neutrino. The key observable, related to the forward-backward asymmetry, is exactly zero for a massless neutrino but becomes non-zero if the neutral lepton is heavy and interacts with Standard Model fields via left-handed operators. For right-handed interactions, this quantity differs significantly from zero even for a massless right-handed neutrino. We demonstrate this explicitly using the example of a pseudoscalar meson decaying into another pseudoscalar meson.}

\vspace{-8mm}
\section{Introduction}
\vspace{-3mm}
\noindent
%%%%%%%%%%%%%%%%%%%%%%%%%%%%%%%%%%%%%%%%
Probing massive neutrinos in low-energy experiments has gained significant interest recently. Various Beyond the Standard Model (BSM) scenarios, including the inverse seesaw mechanism, low-energy leptoquarks, and left-right symmetry, introduce sterile or right-handed (RH) neutral leptons in the MeV–GeV range. These species offer elegant neutrino mass generation mechanisms and potential solutions to discrepancies between Standard Model (SM) predictions and data. In this document, we demonstrate that sub-GeV neutrinos can be effectively tested via semileptonic meson decays. We focus primarily on the decay of
a pseudoscalar meson into another pseudoscalar meson and we define observables that distinguish between left-handed (LH) and right-handed (RH) neutrino couplings, complementing other recent studies in the field. More details can be found in Ref.~\cite{Becirevic:2026tnk}
\vspace{-5mm}
\section{Effective Theory and Observables}
\vspace{-3mm}
\noindent We consider the effective theory for semileptonic decays $d\to u \ell N$, where $N$ can be the SM neutrino or a massive neutral lepton $N_{L,R}$ coupling via LH or RH interactions. The low-energy effective Hamiltonian (LEFT) reads:
\vspace{-3mm}\begin{equation}\label{eq:hamilt}\small
\mathcal{H}_{\text{eff}} = \frac{4 G_F V_{ud}}{\sqrt{2}} 
\sum_{Y, A, B}  C_{AB}^{Y} \mathcal{O}_{AB}^{Y} + \mathrm{h.c.},
\end{equation}\vspace{-7mm}

\noindent with the four-fermion operators defined as:
\vspace{-3mm}\begin{equation}\small
    \label{eq:Ops}
\mathcal{O}_{AB}^{V} = (\bar{u} \gamma^\mu P_A d) (\bar{\ell} \gamma_\mu P_B N),\quad \mathcal{O}_{AB}^{S} = (\bar{u} P_A d) (\bar{\ell} P_B N), \quad
\mathcal{O}_{AB}^{T} = \delta_{AB} (\bar{u} \sigma^{\mu\nu} P_A d) (\bar{\ell} \sigma_{\mu\nu} P_A N). \nonumber
\end{equation}\vspace{-7mm}

\noindent The SM is recovered for $C^{V}_{LL} = 1$ with all other coefficients vanishing. This study focuses on identifying observables sensitive to New Physics ($C^Y_{AB} \neq 0$) in pseudoscalar meson to pseudoscalar meson decays, with extensions to vector mesons discussed in Ref.~\cite{Becirevic:2026tnk}.
%%%%%%%%%%%%%%%%%%%%
%%%%%%%%%%%%%%%%%%%%
\vspace{-5mm}\section{Decay to a pseudoscalar meson}\vspace{-3mm}
\noindent The double differential decay width for $M\to P \ell N$ in the dilepton rest frame is:

\begin{equation}
\small
  \label{eq:ang_coeff_P}
    \frac{d^2 \Gamma^s}{dq^2 d\cos{\theta_\ell}} = a_s(q^2) + b_s(q^2) \cos{\theta_\ell} + c_s(q^2) \cos^2{\theta_\ell}\,,
\end{equation}
\noindent where $s\in \{+,-\}$ denotes the lepton  helicity and $\theta_\ell$ the angle between the lepton and the outgoing meson. The total width $\Gamma = \Gamma^+ + \Gamma^-$ is obtained by integrating over $q^2 \in [(m_\ell + m_N)^2, (m_M - m_P)^2]$. The forward-backward asymmetry for each helicity component isolates the $b_s(q^2)$ function:
\begin{equation}\small
\label{eq:AFB0}
\langle A_{FB}^s \rangle = \frac{1}{\Gamma}  \int_{q_{\text{min}}^2}^{q_{\text{max}}^2}  dq^2 \left( \int_0^1 - \int_{-1}^{0} \right) d\cos\theta_\ell \, \frac{d^2\Gamma^s}{dq^2 d\cos\theta_\ell} = \frac{1}{\Gamma}  \int_{q_{\text{min}}^2}^{q_{\text{max}}^2} dq^2\, b_{s}(q^2) \,.
\end{equation}
\noindent where $\langle A_{FB} \rangle = \langle A_{FB}^- \rangle + \langle A_{FB}^+ \rangle$. Notably, as emphasized in Ref.~\cite{Becirevic:2020rzi}, the SM satisfies:
\vspace{-1mm}\begin{equation}\label{b-SM}\small
b_-(q^2)^\mathrm{SM}=0\,,
\end{equation}\vspace{-6mm}

\noindent implying the SM asymmetry is entirely determined by $b_+(q^2)$.  This observation remains true even in the presence of other (BSM) operators 
in Eq.~\eqref{eq:hamilt}, provided one works with a (massless) SM neutrino.  
For a non-zero $\langle A_{FB}^- \rangle$ the quark currents should either couple to (i) the LH massive neutral lepton, or (ii) to the  massive or \textbf{massless} RH neutral lepton.
In other words, by measuring $\langle A_{FB}^+ \rangle$ and $\langle A_{FB}^- \rangle$, and possibly their $q^2$-shapes, one can obtain a key constraint that can (in)validate 
the scenarios of physics BSM that involve additional neutral lepton species. We reiterate that the previous statement is valid for all the interaction operators given in Eq.\ref{eq:hamilt}.

We have carefully studied all possible situations. In particular, if we restrict ourselves to the SM-like scenario of new physics, in which one only adds a massive neutral lepton (neutrino) without modifying the interaction Hamiltonian, one has $b_-(q^2)\neq 0$, and therefore measuring $\langle A_{FB}^- \rangle \neq 0$ would confirm such scenarios. Even more interesting is the case of a theory in which the SM particle content is extended by a massless RH neutrino. We find that in this situation, $\langle A_{FB}^- \rangle \neq 0$, as well. 
\noindent The presence of additional BSM operators makes the situation somewhat more complicated. Importantly, however, if only $C^{S}_{A,B}$ or $C^{T}_{A,B}$ are allowed to be non-zero, $\langle A_{FB} \rangle$ is not modified at all. A contribution to $\langle A_{FB} \rangle$ can be generated only if we have a SM-like scenario, or if $C^{S}_{A,B}$ or $C^{T}_{A,B}$ are not the only non-zero BSM Wilson coefficients. The resulting shift in $\langle A_{FB}^s \rangle$, however, turns out to be much smaller than in the case in which the Wilson coefficients $C^{V}_{AB}$ are chosen so as to be compatible with the measured branching fraction of a given semileptonic decay mode (within $2\sigma$). 
\noindent In the following, to make the discussion more specific, we consider scenarios in which the BSM contribution arises from couplings to the vector operator in Eq.~\eqref{eq:hamilt}. We stress again that the expressions provided in Ref. \cite{Becirevic:2026tnk} include contributions from all operators present in Eq.~\eqref{eq:hamilt}.

\noindent We thus rewrite Eq.~\eqref{eq:hamilt} as  
$\label{eq:hamilt2}
\mathcal{H}_{\text{eff}} =  \frac{4 G_F V_{ud}}{\sqrt{2}} \,  \sum_{B=L,R}\left(
C^V_{LB}\mathcal{O}_{LB}^V +C^V_{RB} \mathcal{O}^V_{RB} \right)\,,$

\noindent where, again, the SM contribution ($N=\nu$) is recovered by setting $C_{LL}^V=1$. Otherwise the BSM vector operators comprise either the LH  or RH quark current and $B = (L,R)$, depending on the handedness of the neutral lepton field. We now discuss the case in which the interaction involves $N_L$, separately from the case involving $N_R$.
\vspace{-5mm}\subsection{LH neutral lepton}\vspace{-2mm}

\noindent We choose $B=L$ in Eq.~\eqref{eq:hamilt} and compute the coefficients $b_{\pm}(q^2)$ entering Eq.~\eqref{eq:ang_coeff_P} for the semileptonic $M\to P\ell N_L$ decay. We find:
{\small
\begin{align}
   \hspace{-6mm} b_-(q^2) &= -\frac{ {\mathcal{N}}_{N_L}}{q^2}|C^V_{LL}+C^V_{RL}|^2 (m_M^2-m_P^2) \sqrt{\lambda(m_M^2,m_P^2,q^2)}  \times \text{Re}\left[f_+(q^2)f_0^*(q^2)\right] \left[K_{-+}(q^2,m_{N_L})\right]^2 \nonumber \\
    &= |C^V_{LL}+C^V_{RL}|^2 M(m_{N_L},q^2) = M(0,q^2)+|C^V_{LL}+C^V_{RL}|^2 M(m_{N_L},q^2) \nonumber \\
    b_+(q^2) &= \frac{m_M^2-m_P^2}{q^2} \sqrt{\lambda(m_M^2,m_P^2,q^2)} \text{Re}\left[f_+(q^2)f_0^*(q^2)\right] \nonumber  \times \left\{ {\mathcal{N}}_0 \left[K_{+-}(q^2,0)\right]^2 + {\mathcal{N}}_{N_L} |C^V_{LL}+C^V_{RL}|^2 \left[K_{+-}(q^2,m_{N_L})\right]^2\right\} \nonumber \\
    &= P(0,q^2)+|C^V_{LL}+C^V_{RL}|^2 P(m_{N_L},q^2) \label{eq:bL}
\end{align}}
\vspace{-8mm}

\noindent where, in the last line of each equation we separate the SM contribution [$M(0,q^2)$ and $P(0,q^2)$], and we reiterate that $M(0,q^2)=0$, as already emphasized in Eq.~\eqref{b-SM}. We use the standard definition of the triangle function, $\lambda(m_a^2,m_b^2,q^2) =[q^2 - (m_a-m_b)^2 ][q^2 - (m_a+m_b)^2]$, that will be referred to as $\lambda_{ab}$ in the following. 

\noindent The normalization factor then reads,
$\label{eq:norm}\small
	{\mathcal{N}}_{N_L} = \frac{|V_{ud}|^2 G_F^2 }{128 \pi^3 m_M^3} \frac{\sqrt{\lambda_{MP}} \sqrt{\lambda_{\ell N_L}}}{4\, q^2  },$ and ${\mathcal{N}}_{0}$ is trivially obtained from the previous definition after setting $m_{N_L} =0$.

\noindent In the notation of Ref.~\cite{Datta:2022czw}, which we adopt here, the functions $K_{\pm\pm}$ are given by,
\begin{equation}\label{eq:Kpm}\small
	K_{\pm\pm}(q^2,m_{N})\equiv \frac{(E_N+m_N\pm |p_\ell|)(E_\ell+m_\ell \pm |p_\ell |)}{\sqrt{(E_\ell +m_\ell)(E_N+m_N)}}\,,
\end{equation}
\vspace{-5mm}

\noindent in the rest frame of the lepton pair. In the massless neutrino limit we have:
\begin{equation}\small
K_{++}(q^2,0) = 2\sqrt{q^2} \beta_\ell ,\quad K_{+-}(q^2,0) = 2m_\ell \beta_\ell\,, \quad K_{-+}(q^2,0)= K_{--}(q^2,0) = 0\,, 
\end{equation}
where $\beta_\ell = \sqrt{1-m_\ell^2/q^2}$. Note also that $f_{+,0}(q^2)$ in Eq.~\eqref{eq:bL} are the two semileptonic decay form factors 
which encode the nonperturbative QCD dynamics of the hadronic matrix element.

\noindent An important observation is that for our decays we have $q^2>m_\ell^2$, $m_M>m_P$, and since the form factors $f_{0,+}(q^2)$ are real and positive, the function $P(m_{N_L}, q^2)$ in Eq.~\eqref{eq:bL} is strictly positive, while the function $M(m_{N_L},q^2)$ is strictly negative. In other words, if the quark vector current couples to a massive LH neutral lepton field we have $\mathrm{sgn}[b_\pm(q^2)]=\pm 1$. When applied to the neutral current processes to two leptons in the final state, this feature will provide the change of sign of $A_{FB}(q^2)$: it is a combination of the strictly positive $A^+_{FB}(q^2)$ and the strictly negative $A^-_{FB}(q^2)$, and the two functions have different size and (slightly) different $q^2$-shapes.

\vspace{-3mm}\subsection{RH neutral lepton}
\vspace{-1mm}

\noindent We now substitute $B=R$ in Eq.~\eqref{eq:hamilt}, compute the functions $b_{\pm}(q^2)$ for $M\to P\ell N_R$, and find:
\vspace{-2mm}{\small\begin{align}
\label{eq:bR}
    b_-(q^2)&= \frac{ {\mathcal{N}}_{N_R}}{q^2}|C^V_{LR}+C^V_{RR}|^2(m_M^2-m_P^2) \sqrt{\lambda_{MP}}\times \text{Re}\left[f_+(q^2)f_0^*(q^2)\right] \left[K_{+-}(q^2,m_{N_R})\right]^2\\\nonumber
    &=M(0,q^2)+ |C^V_{LR}+C^V_{RR}|^2 P(m_{N_R},q^2)\,,\\\nonumber
    \hspace{-5mm}b_+(q^2)&=\frac{m_M^2-m_P^2}{q^2} \sqrt{\lambda_{MP}} \, \text{Re}\left[f_+(q^2)f_0^*(q^2)\right]\times\left\{  {\mathcal{N}}_0 \left[K_{+-}(q^2,0)\right]^2 + {\mathcal{N}}_{N_R} |C^V_{LR}+C^V_{RR}|^2 
  \left[K_{-+}(q^2,m_{N_R})\right]^2\right\}\\ \nonumber
    &=P(0,q^2)+|C^V_{LR}+C^V_{RR}|^2 M(m_{N_R},q^2)\,,
\end{align}}
\vspace{-8mm}

\noindent where $ {\mathcal{N}}_{N_R}$ is the same as in the LH case after replacing $N_L\to N_R$. In this situation $M(0,q^2)=0$, as explained before, cf. Eq.~\eqref{b-SM}. 
We see that for a massive $m_{N_R}$ the above expressions are simply flipped with respect to the expressions given in Eq.~\eqref{eq:bL}.  
By following the same line of arguments as in the previous subsection, and by keeping in mind that the two form factors are real and positive, we easily deduce 
that, this time, the function $b_-(q^2)$ is strictly positive. Therefore, 
a measurement of the sign of $\langle A_{FB}^-\rangle$ could help us distinguishing between the coupling to the LH from that to the RH neutral lepton. 
If the quark vector current couples to a massive LH neutrino then one has $\langle A_{FB}^-\rangle < 0$, while the coupling to the RH one results in $\langle A_{FB}^-\rangle > 0$.  
This information can be quite useful, provided $\langle A_{FB}^-\rangle$ can be measured. 
We stress once again that in the massless case, $b_-(q^2)=0$, for the LH neutrino, while it is non-zero and positive for the RH one. 
\vspace{-4mm}\section{Illustrative example}
\vspace{-2mm}
\noindent To illustrate our results, we focus on the decay $B\to D\tau N$. This mode is of particular interest as measurements of $R_{D^{(*)}}= \mathcal{B}(B \to D^{(*)} \tau\bar\nu)/\mathcal{B}(B \to D^{(*)} \mu\bar\nu)$ show a $3\div 4\sigma$ discrepancy with the SM~\cite{HeavyFlavorAveragingGroupHFLAV:2024ctg}, potentially signaling lepton flavor universality violation. In the literature, this hint has been adressed by invoking a right-handed neutrino ($N_R$) with a mass in the MeV--GeV range~\cite{Becirevic:2024iyi,Azatov:2018kzb}. We use experimental inputs from Ref.~\cite{ParticleDataGroup:2024cfk} and form factors from Ref.~\cite{FlavourLatticeAveragingGroupFLAG:2024oxs}. Extracting bounds on the Wilson coefficients $C^V_{AB}$ to $2\sigma$ from experimental decay widths, we test the sensitivity of $\langle A_{FB}^\pm\rangle$ to the presence of $N_L$ or $N_R$.

\begin{SCfigure}[0.5][t!] % [Relative width of caption][placement]
  \centering \vspace{-9mm}
  \includegraphics[width=0.55\linewidth]{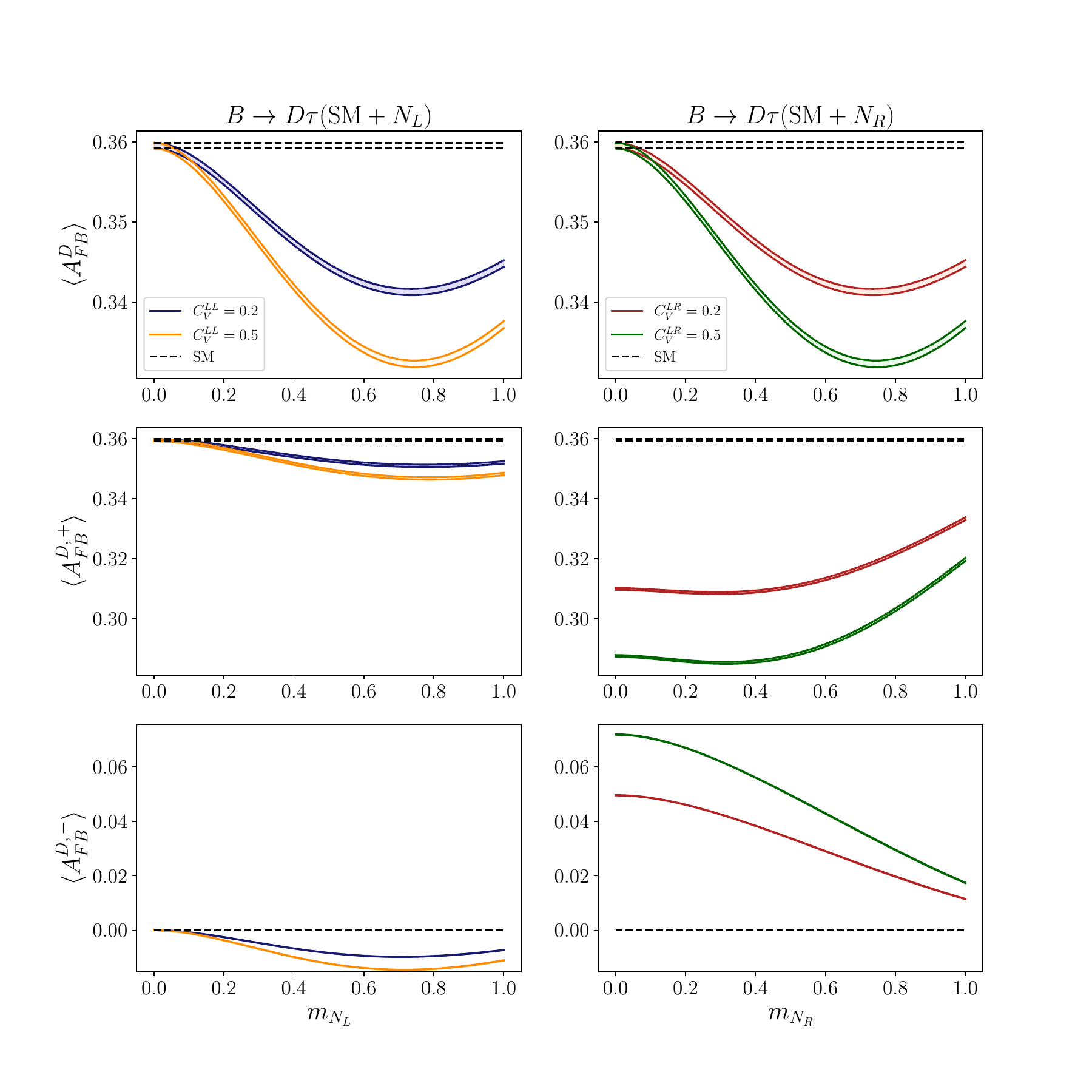}
  \caption{\small \sl Two components of the forward--backward asymmetry of $B \to D \tau \text{`inv'}$. The left (right) plots show $\langle A_{FB}^\pm\rangle$ as a function of the mass of $N_L$ ($N_R$).}
  \label{fig:Afb}
  \vspace{-2mm}
\end{SCfigure}

We show the integrated $\langle A_{FB}^\pm\rangle$ in Fig.~\ref{fig:Afb} for the case of the SM being extended by $N_L$, separately from the case of $N_R$. We see that for a massless $N_L$ one recovers the SM case, but in the case of a massless $N_R$ the effect is rather important and both $\langle A_{FB}^+\rangle$ and $\langle A_{FB}^-\rangle$ are significantly different from the SM case. Note that we chose two values of the Wilson coefficients $C_V^{LL}$ ($C_V^{LR}$) that are consistent with the experimental value of $\mathcal{B}(B\to D\tau \text{`inv'})$.
\noindent As for the experimental feasibility of such a study, we should mention that in Ref.~\cite{Belle:2017ilt} it was shown that $\Gamma^+$ and $\Gamma^-$ could be successfully separated. A new complexity is that in addition to that separation of events, an angular analysis should be done on each subset in order to expose the effect we are after in this paper.
%%%%%%%%%%%%%%%%%%%%%%%%%%%%%%%%%%%%%%%%
\vspace{-4mm}\section{Summary}\label{sec:summary}
\vspace{-3mm} % Pulls section header up

We have shown that semileptonic decays with separated lepton polarization provide a powerful probe for massive $\mathcal{O}(\text{MeV--GeV})$ neutrinos~\cite{Becirevic:2026tnk}. The forward–backward asymmetry, specifically the component $\langle A_{FB}^-\rangle$, serves as a smoking-gun observable: it is exactly zero in the Standard Model but becomes non-zero if the neutral lepton is massive or possesses right-handed interactions~\cite{Becirevic:2020rzi}. Crucially, the sign of $\langle A_{FB}^-\rangle$ allows for a clear distinction between couplings to left-handed and right-handed neutral leptons in the case of vector interactions. While our primary focus was on $M \to P \ell N$ transitions, these conclusions extend to vector meson decays and general effective theories including scalar and tensor operators~\cite{Becirevic:2026tnk}.

\vspace{-3mm}\section*{Acknowledgments}
\vspace{-2mm}
\noindent S.F., N.K. and L.P. acknowledge financial support from the Slovenian Research Agency (research core funding No. P1-0035, N1-0321 and N1-0407).
This project has received support from the European Union’s Horizon 2020 research and innovation program under the Marie Sklodowska-Curie grant agreement No 860881-HIDDeN.
%\newpage

\vspace{-4mm}\section*{References}\vspace{-3mm}
\small\bibliography{chevallier}

%%% manually generated bibliography
%\begin{thebibliography}{99}
%\bibitem{ja}C Jarlskog in {\em CP Violation}, ed. C Jarlskog
%(World Scientific, Singapore, 1988).
%\bibitem{ma}L. Maiani, \Journal{\PLB}{62}{183}{1976}.
%\bibitem{bu}J.D. Bjorken and I. Dunietz, \Journal{\PRD}{36}{2109}{1987}.
%\bibitem{bd}C.D. Buchanan {\it et al}, \Journal{\PRD}{45}{4088}{1992}.
%\end{thebibliography}

\end{document}